\documentstyle{article}\begin{document}
\title{ Response of a canonical ensemble of quantum oscillators to  a random metric}

\author{Z. Haba \\
Institute of Theoretical Physics, University of Wroclaw,\\ 50-204
Wroclaw, Plac Maxa Borna 9, Poland\\
email:zbigniew.haba@uwr.edu.pl} \maketitle
\begin{abstract} We calculate the susceptibility of a canonical ensemble
of quantum oscillators to the singular random metric. If the
covariance of the metric is $\vert {\bf x}-{\bf x}^{\prime}\vert
^{-4\alpha}$ ($0<\alpha<\frac{1}{2}$)then the expansion of the
partition function in powers of the temperature involves
non-integer indices .\end{abstract}
\maketitle\section{Introduction} It is known that the dynamics in
an irregular domain can be chaotic and conversely the chaotic
motion can have a fractal attractor \cite{chaos}\cite{turbulence}.
  Diffusion in fractal domains exhibits their fractal dimension
which in general is not a natural
number\cite{wil}\cite{dunne}\cite{carlip}\cite{mandel}.
Thermodynamics of a canonical ensemble of particles  in an
irregular domain depends on the spectral dimension of the domain
\cite{dunne2}. In such a case thermodynamic properties (e.g.
critical indices) may be functions of a non-integer dimension. For
some time fractal geometry has been associated with quantum
gravity \cite{carlip}\cite{jurk}\cite{horava}.Quantum gravity can
be expressed as a random geometry.
  It is believed that quantum gravity leads
to a fractal geometry by a "foamy" behaviour of the metric at
short distances \cite{smolin}. Such irregular shapes of random
figures have been at the basis of the fractal geometry
\cite{mandel}.It has been suggested that irregular metric at the
Planck scale can modify the short distance behaviour of quantum
fields at short distances \cite{carlip}\cite{smolin}\cite{haba}.

  In this paper we study quantum oscillators in a random
singular metric ( or random position dependent mass). We define
the susceptibility to the metric which has an expansion in powers
of the inverse temperature $\beta$ if the metric is a regular
random field. We show that if the random metric is singular then
the susceptibility has an expansion in non-integer powers of
$\beta$. It is known that an ensemble of oscillators can serve as
an approximation to field theory. The quantum statistical
mechanics of oscillators will resemble the quantum field theory at
finite temperature.

We consider a  Hamiltonian $H_{\gamma}$ perturbed around the free
theory(harmonic oscillators)
\begin{equation} H_{\gamma}=H+\gamma H_{1}.
\end{equation}
The statistical expectation value of an observable ${\cal A}$ at
the temperature $\beta^{-1}=k_{B}T$ (where $k_{B}$ is the Boltzman
constant)  is defined by

\begin{equation}
<{\cal A}>_{\beta}= Z_{\gamma}^{-1} Tr\Big(\exp(-\beta
H_{\gamma}){\cal A}\Big),
\end{equation}where the partition function
\begin{equation}
Z_{\gamma}=Tr\Big(\exp(-\beta H_{\gamma})\Big).
\end{equation}
For a small $\gamma$ we have the expansion (till the first order
in $\gamma$)
\begin{displaymath}
Z_{\gamma}=Z_{0}-\gamma\int_{0}^{\beta}dsTr\Big(\exp(-\beta
H)\exp(sH)H_{1}\exp(-sH)\Big).\end{displaymath}

In terms of the partition function we can define other
thermodynamic functions as ,e.g.,the internal energy $U$
\begin{equation}
U_{\gamma}=-\partial_{\beta}\ln Z_{\gamma}.
\end{equation}
The susceptibility to $H_{1}$ can be defined as \begin{equation}
\chi_{\beta}=\partial_{\gamma}(Z_{\gamma})_{\vert \gamma=0}.
\end{equation}We can use the
expansion
\begin{displaymath}
\exp(sH)H_{1}\exp(-sH)=H_{1}+s[H,H_{1}]+....
\end{displaymath}
in eq.(5) to see that for a small $\beta$  \begin{equation}
\chi_{\beta}=\partial_{\gamma}(Z_{\gamma})_{\vert \gamma=0}=-\beta
Tr\Big(\exp(-\beta H)H_{1}\Big).
\end{equation}

For $H_{0}$ of the harmonic oscillator with the  frequency
$\omega$ ( the ground state energy subtracted) we have
\begin{equation} \ln Z_{0}=-\ln(1-\exp(-\beta \omega))
\end{equation}and
\begin{equation}
U=-\partial_{\beta}\ln Z_{0}=\omega(\exp(\beta
\omega)-1)^{-1}\simeq \beta^{-1}
\end{equation}
for a small $\beta$ (high temperature).

If the oscillators are the modes of the electromagnetic field in a
cavity then $\omega =\vert {\bf k} \vert c$ where $c$ is the
velocity of light and ${\bf k}$ is the wave vector in the cavity.
In such a case in eqs.(7)-(8) we have a sum over modes. The
modified thermodynamics \cite{dunne2} comes from the modified
distribution of modes in cavities with fractal geometry.

We consider Hamiltonians $H$ such that its similarity
transformation
\begin{equation}
\hat{H}=\Omega^{-1} H\Omega
\end{equation}
gives  a generator of a diffusion process ($\exp(-\beta \hat{H})$
is a Markov semigroup).

According to our assumption
\begin{equation}
(\exp(-\beta \hat{H})\psi)(\xi)=E[\psi(\xi_{\beta}(\xi))],
\end{equation} where $\xi$ are the coordinates of the oscillators, $\xi_{\beta}(\xi)$ is a Markov
process starting from $\xi$ and the expectation value $E[...]$ is
over the paths of the process. In our models we consider
$\xi=({\bf x},X)\in R^{n+d}$ and assume that $H_{1}=V(P)$ (a
function of momentum). Then, according to eq.(6)
\begin{equation}
\begin{array}{l}
Tr\Big(\exp(-\beta
\hat{H})V(P)\Big)\cr=(2\pi)^{-d}E\Big[\delta({\bf x}_{\beta}({\bf
x})-{\bf x}) \exp(iP(X_{\beta}-Y)) <Y\vert V(P)\vert X>\Big]d{\bf
x}dXdYdP,
\end{array}\end{equation}where
\begin{equation}
<Y\vert V(P)\vert X>=(2\pi)^{-d}\int dK\exp(iK(Y-X))V(K).
\end{equation}
So that\begin{equation}
\begin{array}{l}
Tr\Big(\exp(-\beta \hat{H})V(P)\Big)=(2\pi)^{-d}\int
E\Big[\delta({\bf x}_{\beta}({\bf x})-{\bf
x})\exp(iP(X_{\beta}-X))\Big] V(P)d{\bf x}dXdP.
\end{array}\end{equation}
\section{Statistical  mechanics of oscillators}
Let us consider in $R^{n+d}$ the coordinates $\xi^{A}$ and a
diffusion  operator of the form (sum over repeated indices)
\begin{equation}\begin{array}{l}
\hat{H}=-\frac{\sigma^{2}}{2}g^{AB}\partial_{A}\partial_{B}+
\omega_{A}
\xi^{A}\partial_{A}+\frac{\omega_{D}}{2}g^{CD}\partial_{C}g_{AB}\xi^{A}\xi^{B}\partial_{D}.\end{array}\end{equation}
By means of
\begin{displaymath}
\Omega=\exp\Big(-\frac{\omega_{A}}{2\sigma}g_{AB}\xi^{A}\xi^{B}\Big)
\end{displaymath}
we obtain the Hamiltonian $H$ of eq.(9)
\begin{equation}\begin{array}{l}
H=-\frac{\sigma^{2}}{2}g^{AB}\partial_{A}\partial_{B}
+\frac{\omega_{A}^{2}}{2}g_{AB}\xi^{A}\xi^{B}\cr+\sigma\frac{\omega_{C}}{8}g^{CD}\partial_{C}g_{RM}\xi^{M}\xi^{R}
\partial_{D}g_{AB}\xi^{A}\xi^{B}-\frac{1}{2}\sum_{A}\omega_{A}.
\end{array}\end{equation}We divide the coordinates $\xi=({\bf x},X)$ into two classes ${\bf x}$ and $X$ where ${\bf
x}\in R^{n}$ and $ X\in R^{d}$. In order to simplify the model we
assume that only the $X$ coordinates are coupled to the random
metric (or a random mass).   So,  $(g^{AB})=(1,g^{\mu\nu}$) and in
$g^{\mu\nu}({\bf x})$ the dependence on $X$ is negligible (the
coordinates $x_{j}$ of ${\bf x}$ have Latin indices $j=1,...,n$
and the coordinates of $X$ the Greek indices $\mu=n+1,..,n+d $).
We can imagine a random mass distribution (producing the metric)
which depends only on some coordinates. To make the model simple
we assume $\omega_{j}=\nu $ and $\omega_{\mu}=\omega$. If
$\omega_{A}$ are the modes of a massless field in a cavity which
is a rectangular box of sides $L_{A}$ then $\omega_{A}=c\frac{2\pi
n_{A}}{L_{A}}$ where $n_{A}$ are integers. We could arrange the
model so that $\omega_{j}$ are small and the non-linear terms in
eq.(14) with $\omega_{j}=\nu$ are negligible. Then, $\hat{H}$ is
of the form
\begin{equation}\begin{array}{l}
\hat{H}=-\frac{\sigma^{2}}{2}\nabla_{{\bf x}}^{2}+\nu{\bf
x}\nabla_{\bf x}+\frac{\omega}{2}\nabla_{\bf
x}g_{\mu\nu}X^{\mu}X^{\nu}\nabla_{\bf
x}-\frac{\sigma^{2}}{2}g^{\mu\nu}({\bf
x})\partial_{\mu}\partial_{\nu}+ \omega
X^{\mu}\partial_{\mu},\end{array}\end{equation} where
\begin{displaymath}
\partial_{\mu}=\frac{\partial}{\partial X^{\mu}}.
\end{displaymath}
With
\begin{equation}
\Omega=\exp\Big(-\frac{1}{2}\nu\sigma^{-1}{\bf
x}^{2}-\frac{1}{2}\omega\sigma^{-1}g_{\mu\nu}X^{\mu}X^{\nu}\Big)
\end{equation}
the similarity transformation (9) gives
\begin{equation}
\begin{array}{l} H=-\frac{\sigma^{2}}{2}\nabla_{{\bf
x}}^{2}+\frac{1}{2}\nu^{2}{\bf x}^{2}
+\sigma\frac{\omega}{8}\nabla_{\bf
x}g_{\mu\nu}X^{\mu}X^{\nu}\nabla_{\bf
x}g_{\sigma\rho}X^{\sigma}X^{\rho}\cr
 -\frac{\sigma^{2}}{2}g^{\mu\nu}({\bf
x})\partial_{\mu}\partial_{\nu}
+\frac{1}{2}\omega^{2}g_{\mu\nu}X^{\mu}X^{\nu}-\frac{d}{2}\omega-\frac{n}{2}\nu.\end{array}\end{equation}
$ \exp(-t\hat{H})$ can be expressed (according to eq.(10)) by the
solution of the stochastic equations \cite{ikeda}
\begin{equation}
d{\bf x}_{t}=-\nu {\bf x}_{t}dt -\frac{\omega}{2}\nabla_{\bf
x}g_{\mu\nu}X^{\mu}X^{\nu}+\sigma d{\bf b}_{t},
\end{equation}
\begin{equation}
dX^{\mu}_{t}=-\omega X^{\mu}_{t}dt+\sigma e^{\mu}_{a}({\bf
x}_{t})dB^{a}_{t},
\end{equation}
where we expressed the metric $g$ by vierbeins (tetrads)  $e$
\begin{equation}
g^{\mu\nu}=e^{\mu}_{a}e^{\nu}_{a}.
\end{equation}  $({\bf b}_{t}, B_{t})$ is the Brownian motion on
$R^{n+d}$ ,i.e., the Gaussian process with mean zero and the
covariance
\begin{displaymath}
E[b^{j}_{t}b_{s}^{l}]=min(t,s)\delta^{jl}
\end{displaymath}
 (and similarly for $B_{t}$). In order to take the expectation value over the metric
 we need an explicit solution of eq.(20). For ${\bf x}_{t}$ we require only some estimates
 on the behaviour in $t$ .
  However, for a simplicity of the arguments
we neglect the non-linear term in eq.(19) (we assume that
$e^{\mu}_{a}=\delta^{\mu}_{a}+\kappa\epsilon^{\mu}_{a}$, where
$\kappa$ is a small parameter, then $\nabla g\simeq \kappa$).
After a negligence of the non-linear term the solution of eq.(19)
with the initial condition ${\bf x}$ is
\begin{equation} {\bf x}_{t}=\exp(-\nu t){\bf
x}+\sigma\int_{0}^{t}\exp(-\nu(t-s))d{\bf b}_{s}.
\end{equation}
The solution of eq.(20) reads
\begin{equation} X^{\mu}_{t}=\exp(-\omega
t)X^{\mu}+\sigma\int_{0}^{t}\exp(-\omega(t-s))e^{\mu}_{a}({\bf
x}_{s})dB_{s}^{a}.
\end{equation}
The kernel $K$ of $\exp(-\beta\hat{H})$  can be expressed by means
of the Fourier transform
\begin{equation}\begin{array}{l}
K_{\beta}({\bf x},X;{\bf y},Y)=(2\pi)^{-d}\int dP
E\Big[\delta({\bf x}_{\beta}({\bf x})-{\bf y})\exp\Big(
iP(\exp(-\beta\omega)X\cr+\sigma\int_{0}^{\beta}\exp(-\omega(\beta-s))e_{a}({\bf
x}_{s})dB^{a}-Y)\Big)\Big].
\end{array}\end{equation}

\section{Random metric}
We assume that $e^{\mu}_{a}$ are Gaussian variables with the mean
$\delta^{\mu}_{a}$ \begin{equation}
e^{\mu}_{a}=\delta^{\mu}_{a}+\kappa\epsilon^{\mu}_{a}
\end{equation}
where
\begin{equation}
<\epsilon^{\mu}_{a}({\bf x})\epsilon^{\nu}_{c}({\bf
x}^{\prime})>=\delta^{\mu\nu}_{ac}G({\bf x}-{\bf x}^{\prime}).
\end{equation}
We calculate the Gaussian integral in eqs.(13) and (24)(with
$\delta^{\mu\nu}_{ac}=\delta^{\mu\nu}\delta_{ac}$ chosen for
simplicity)
\begin{equation}\begin{array}{l}
<\exp(iP(\exp(-\omega
\beta)X+\sigma\int_{0}^{\beta}\exp(-\omega(\beta-s))e_{a}({\bf
x}_{s})dB^{a}))>\cr=\exp\Big(iP\exp(-\omega\beta)X-\frac{1}{2}<(PQ_{\beta})^{2}>
+i\sigma
P_{\mu}\int_{0}^{\beta}\exp(-\omega(\beta-s))dB_{s}^{\mu}\Big))\cr
=\exp\Big(-\frac{1}{2}\kappa^{2}P_{\mu}P_{\nu}\int_{0}^{\beta}\int_{0}^{\beta}\exp(-\omega(\beta-s))
\exp(-\omega(\beta-s^{\prime})) G({\bf x}_{s}-{\bf
x}_{s^{\prime}})dB^{\mu}_{s}dB^{\nu}_{s^{\prime}}\cr+iP\exp(-\omega
\beta)X+i\sigma
P_{\mu}\int_{0}^{\beta}\exp(-\omega(\beta-s))dB_{s}^{\mu}\Big).
\end{array}\end{equation}
In eq.(27) we used the formula for  Gaussian expectation value of
$\epsilon$

\begin{equation}
<\exp(iP_{\mu}Q_{\beta}^{\mu})>=\exp(-\frac{1}{2}<(P_{\mu}Q_{\beta}^{\mu})^{2}>)\end{equation}
with\begin{equation}
P_{\mu}Q^{\mu}_{t}=P_{\mu}\sigma\int_{0}^{t}\exp(-\omega(t-s))\epsilon^{\mu}_{a}({\bf
x}_{s})dB_{s}^{a}\equiv F_{t}.
\end{equation}
This expression can be written in a different way using the Ito
calculus \cite{ikeda}\cite{simon}
\begin{equation}
\int dF_{t}^{2}=2\int F_{t}dF_{t}+\int dF_{t}dF_{t},
\end{equation}
where we have (here $P^{2}=P_{\mu}P^{\mu}$)
\begin{equation}
dF_{t}dF_{t}=d\kappa^{2}G({\bf 0}) P^{2}dt.
\end{equation}Next, we consider a singular covariance G. For this purpose
at the beginning we treat $\epsilon^{\mu}_{a}$ as a regularized
random field.  Then, we remove the regularization. In order to
make $H\psi$  a well-defined random field we need the normal
ordering of $H$ with
\begin{displaymath}
:g^{\mu\nu}:=:e^{\mu}_{a}({\bf x})e^{\nu}_{a}({\bf
x}):=e^{\mu}_{a}({\bf x})e^{\nu}_{a}({\bf
x})-\kappa^{2}<\epsilon^{\mu}_{a}({\bf x})\epsilon^{\nu}_{a}({\bf
x})>.
\end{displaymath}
The normal ordering in the exponential of eq.(27) removes the
second term $ d\kappa^{2}G({\bf 0}) P^{2}t$ on the rhs of eq.(30)
whereas the first term there (i.e., $\int_{0}^{\beta}F_{s}dF_{s})$
becomes a time-ordered integral (renormalization of such
expressions appear also in QED \cite{betz}\cite{albeverio}). Owing
to the normal ordering in eq.(27)
\begin{equation}
<(PQ_{t})^{2}>\rightarrow <( PQ_{t})^{2}>-t P^{2}\kappa G(0)d
\end{equation}
and because of the subtraction  (32) we can define the action of
$\exp(-tH)$ upon a test function $\psi$ (decaying fast in the
momentum space) so that $\exp(-Ht)\psi$ is a well-defined random
field. After the averaging over the translation invariant random
field $e^{\mu}_{a}({\bf x})$ and the renormalization (32) we can
write $<\exp (-\beta :H:)\psi>$ in terms of Fourier transforms as
\begin{equation}\begin{array}{l}(<\exp (-\beta :H:)>\psi)({\bf x},X)=\cr
E\Big[\delta({\bf x}_{\beta}-{\bf y})
\exp\Big(-\kappa^{2}P_{\mu}P_{\nu}\int_{0}^{\beta}\exp(-\omega(\beta-s))
dB^{\mu}_{s}\int_{0}^{s}dB^{\nu}_{s^\prime}\exp(-\omega(\beta-s^{\prime}))
G({\bf x}_{s}-{\bf x}_{s^{\prime}})\cr+i\sigma
P_{\mu}\int_{0}^{\beta}\exp(-\omega(\beta-s))dB_{s}^{\mu} +
iP\exp(-\omega\beta)X)\Big)\Big]\psi({\bf y},P)dPd{\bf y}.
\end{array}\end{equation}
At $\kappa=0$  we have
\begin{equation}\begin{array}{l}
Z_{0}=Tr(\exp(-\beta H_{0}))=\int d{\bf x}dX<\exp (-\beta
H_{0})>({\bf x},X;{\bf x},X)=\cr (2\pi)^{-d}\int dPd{\bf x}dX
E\Big[\delta({\bf x}_{\beta}-{\bf
x})\exp(iP((\exp(-\omega\beta)-1)X +i\sigma
P\int_{0}^{\beta}\exp(-\omega(\beta-s))dB_{s}))\Big]\cr =\int
K^{(0)}_{\beta}({\bf x},X;{\bf x},X)d{\bf x}dX,
\end{array}\end{equation}
where $H^{(0)}$ is the Hamiltonian of uncoupled harmonic
oscillators and $K^{(0)}$ is the  well-known Mehler kernel of the
harmonic oscillator. So, at $\kappa=0$ we obtain the formula (7).
We are interested in the $\kappa$-term as a perturbation resulting
from an interaction with a random metric.
 In eq.(33) we apply the identity for $
B_{s}$ and ${\bf b}_{s}$ (in the sense that both sides have the
same probability law)
\begin{equation}
B_{s}=\sqrt{\lambda} B_{\frac{s}{\lambda}}.
\end{equation}After rescaling
\begin{equation} {\bf x}_{s^{\prime}}=\exp(-\nu
s^{\prime}){\bf
x}+\sqrt{\lambda}\int_{0}^{\frac{s^{\prime}}{\lambda}}\exp\Big(-\lambda\nu(\frac{s^{\prime}}{\lambda}-\tau)\Big)d{\bf
b}_{\tau}.
\end{equation}
It is easy to see that for a small t in eq.(22) (set
$\lambda=s^{\prime}$ in eq.(36))
\begin{equation}
{\bf x}_{t}={\bf x}+\sqrt{t}{\bf q}_{t},
\end{equation}
where ${\bf q}_{t}\simeq {\bf a}+ {\bf c}\sqrt{t} +...$ with
$\vert {\bf a}\vert>0$ for a small $t$.
 This scaling behaviour is all what we need to assume about solutions
of eq.(19) with a random metric $g$ of eq.(26). In eq.(33) we
rescale the Brownian motion $B_{s^{\prime}}$ and ${\bf
x}_{s^{\prime}}$ as in (36) (with $\lambda=s $) and subsequently
$B_{s}$ and ${\bf x}_{s}$ with $\lambda=\beta$. After such a
change of variables the integral of the $\kappa$-dependent part in
the formula (33) reads
\begin{equation}\begin{array}{l}
\exp\Big(-\frac{1}{2}\kappa^{2}\beta
P_{\mu}P_{\nu}\int_{0}^{1}\exp(-\beta\omega(1-s))dB^{\mu}_{s}\int_{0}^{s}
\exp(-\beta\omega(1-s^{\prime}))G({\bf x}_{s}-{\bf
x}_{s^{\prime}})dB^{\nu}_{s^{\prime}}\Big),
\end{array}\end{equation}where
$0\leq s^{\prime}\leq s\leq 1$, ${\bf x}_{s}={\bf
x}+\sqrt{\beta}{\bf q}_{s}$ and ${\bf x}_{s^{\prime}}={\bf
x}+\sqrt{\beta}\tilde{{\bf q}}_{s^{\prime}}$ where ${\bf
q}_{s}\neq 0$ and $\tilde{{\bf q}}_{s}\neq 0$ at $\beta=0$. We
consider the covariance (the upper bound on $\alpha>0$ will be
discussed at the end of this section)
\begin{equation}
G({\bf x}-{\bf x}^{\prime})=\vert {\bf x}-{\bf
x}^{\prime}\vert^{-2\alpha}.
\end{equation}
 Then, from eq.(38) for a small $\beta$
\begin{equation}
G({\bf x}_{s}-{\bf
x}_{s^{\prime}})=\beta^{-\alpha}g(\beta,s,s^{\prime}),
\end{equation}
where $s^{\prime}\leq s\in[0,1]$ and $g\simeq A+C\beta$ (with
$A>0$) for a small $\beta$. Hence,eq.(33) is of the form
\begin{equation}\begin{array}{l}
E\Big[\exp\Big(iP((\exp(-\omega
\beta)-1)X-\int_{0}^{\beta}\exp(-\omega(\beta-s))dB)\Big)
\cr\times\exp\Big(-\kappa^{2}\beta^{1-\alpha}P_{\mu}P_{\nu}f^{\mu\nu}(\beta)\Big)\Big],
\end{array}\end{equation}
where $\vert f^{\mu\nu}(\beta)P_{\mu}P_{\nu}\vert$ is bounded from
below by a constant.

We can now estimate the behaviour of the partition function in the
metric field (26) ( note that $:g^{\mu\nu}:- <:g^{\mu\nu}:>$ has
the covariance $\vert {\bf x}-{\bf x}^{\prime}\vert^{-4\alpha}$ if
$\epsilon^{\mu}_{a}$ has the covariance (39)). We have
\begin{equation}
g^{\mu\nu}=\delta^{\mu\nu}+2\kappa\epsilon^{\mu}_{\nu}+\kappa^{2}\epsilon^{\mu}_{a}\epsilon^{\nu}_{a}.
\end{equation}
The Hamiltonian (16) is of the form $H=H_{0}+\kappa {\cal
H}_{1}+\kappa^{2}{\cal H}_{2}$.From eq.(27) we can see that the
contribution to the partition function $Z$ is of order
$\kappa^{2}$. We consider $H_{\gamma}=H+\gamma V(P)$ then in the
approximation (6) \begin{equation}
\partial_{\gamma}Z_{\gamma}=Tr(\exp(-\beta H)V(P)).
\end{equation}
The anomalous (fractional)  dependence $\beta^{1-\alpha}$ in
eq.(41) of the partition function is a characteristic of the
coupling to a singular metric field. We can  calculate the
$\kappa^{2}$-derivative of the susceptibility (6) to the metric
for small $\beta$ (high temperature)

\begin{equation}\begin{array}{l}
\partial_{\kappa^{2}}\partial_{\gamma}Z_{\vert\kappa=\gamma=0}=
\beta^{1-\alpha} E\Big[P_{\mu}P_{\nu}f^{\mu\nu}(\beta) \delta({\bf
x}_{\beta}-{\bf x}) \cr\exp(iP(\exp(-\omega \beta)-1)X+i\sigma
P\int_{0}^{\beta}\exp(-\omega(\beta-s))dB_{s}) )\Big]\beta
V(P)d{\bf x}dXdP
\end{array}\end{equation}and \begin{equation}\begin{array}{l}
\partial^{2}_{\kappa^{2}}\partial_{\gamma}Z_{\vert\kappa=\gamma=0}=
\beta^{2-2\alpha}E\Big[\Big(P_{\mu}P_{\nu}f^{\mu\nu}(\beta)\Big)^{2}\delta({\bf
x}_{\beta}-{\bf x}) \cr \exp(iP(\exp(-\omega \beta)-1)X+i\sigma
P\int_{0}^{\beta}\exp(-\omega(\beta-s))dB_{s})        )\Big] \beta
V(P)d{\bf x}dXdP.
\end{array}\end{equation}
For a  small $\kappa$ and small $\beta$ the partition function has
the expansion in non-integer powers of $\beta$
\begin{equation}
\partial_{\gamma}Z_{\kappa^{2}}=\partial_{\gamma}Z_{0}+\kappa^{2}\chi_{1}\beta\beta^{1-\alpha}
+\kappa^{4}\chi_{2}\beta\beta^{2-2\alpha}+....
\end{equation}with certain constants $\chi_{1}$ and $\chi_{2}$.

We still have to estimate the integrals in eqs.(43)-(46). The
integral (43) is expressed by kernels in eq.(13). The expectation
value  in eq.(13) after the renormalization (32) involves the
time-ordered stochastic integral in eq.(33) which fails to be
positive definite. Hence, if the expectation value in eq.(43) is
to be finite $V(P)$ must decrease faster than $\exp(-RP^{2})$ for
any $R$ (the derivatives in eqs.(44)-(46) impose milder
requirements on the decay of $V(P)$ for a large $P$). There is
still the problem of the convergence of the stochastic integrals
in the definition of $f_{\mu\nu}P^{\mu}P^{\nu}$. This stochastic
integral is of the form
\begin{equation}
P_{\mu}P_{\nu}\int_{0}^{\beta}
dB^{\mu}_{s}\int_{0}^{s}dB^{\nu}_{s^{\prime}}G({\bf x}_{s}-{\bf
x}_{s^{\prime}}).
\end{equation}
The stochastic integrals in eq. (47) can be estimated by ordinary
integrals using the formula \cite{gikhman}(better estimates on the
multiple stochastic integrals (33) and (47) can be obtained using
the results of ref.\cite{carlen})
\begin{equation}
E\Big[\Big(\int FdB_{s}\Big)^{2k}\Big]\leq C_{k}E\Big[\int
F^{2k}ds\Big]
\end{equation} with certain constants $C_{k}$.
The rhs of eq.(48) involves the Ornstein-Uhlenbeck process (22)
\cite{chandra} whose transition function is expressed by the
Mehler formula. In calculations in eq.(48) for small $\beta$ we
can approximate  the Ornstein-Uhlenbeck process ${\bf x}_{s}$ by
the Brownian motion with the transition function $p(s,{\bf
x})=(2\pi s)^{-\frac{n}{2}}\exp(-\frac{\vert{\bf
x}\vert^{2}}{2s})$. Then, the integral in eq.(47) can be estimated
by (use eq.(48) twice with $k=1$ )
\begin{equation}
\int dsds^{\prime}\int d{\bf x}p(s-s^{\prime},{\bf x})\vert {\bf
x}\vert^{-4\alpha}<\infty
\end{equation}
if $2\alpha <1$. The expansion (46) must be terminated at the
$k$-th order if $2k\alpha>1$  because the rhs of the estimate (48)
is infinite.

\section{The outlook}
Some approximate calculations \cite{jurk}\cite{horava}\cite{haba}
indicate that the singularity of the quantum gravitational field
at small distances can be different than the canonical one  which
in $n$ dimensions is of the form $\vert {\bf x}-{\bf
x}^{\prime}\vert^{-n+2} $ ( where $n=4$ corresponds to
$\alpha=\frac{1}{2}$ in eq.(39)). The fractal dimensionality of
the physical  space-time has been discussed in
\cite{brazil}\cite{amelia} on the basis of the Cosmic Microwave
Background measurements. Some limits on the deviation of the
observational space-time dimension from the physical four
dimensions have been obtained. The effect of quantum gravity could
be observed either at small distances or at high energies (which
in cosmology are connected with high temperatures). The
non-integer indices in the expansion of the partition function  in
eqs.(44)-(46) could indicate the relevance of quantum gravity for
some extremal processes in astrophysics ( which possibly could be
tested on the quantum level in gravitational wave interferometers
\cite{wilczek}). The model of a random mass distribution which
according to eqs.(14) and (23) is equivalent to a random
diffusivity is of interest in condensed matter physics
\cite{complex}\cite{comp}. An anomalous behaviour of the partition
function (46) or other thermodynamic functions of complex systems
(e.g. molecules or crystals) could be an indication of the random
mass or random metric present in these systems.

\end{document}